\documentclass[aps,twocolumn,aps,superscriptaddress,amssymb,showpacs ]{revtex4}
\usepackage{ulem}
\usepackage{graphicx}
\usepackage{bm}
\usepackage{epsfig}
\usepackage{color}

\usepackage{amsmath}
\usepackage{amssymb}
\usepackage{cancel}
\usepackage{ulem}
\usepackage[usenames,dvipsnames]{xcolor}

\newcommand{\imat}{{\mathrm{i}}}

\newcommand\jldsout[1]{\bgroup\markoverwith{\textcolor{red}{\rule[0.5ex]{2pt}{0.4pt}}} \ULon{#1} }

\begin{document}

\title{Quantum Goos-H\"{a}nchen shift and tunneling transmission at a curved step potential}

\author{Soo-Young Lee}
\email{pmzsyl@gmail.com}
\address{School of Electronics Engineering, Kyungpook National University, Daegu, 702-701, South Korea}
\address{Max-Planck-Institut f\"ur Physik komplexer Systeme,
  N\"othnitzer Str. 38, D01187 Dresden, Germany}

\author{J\'{e}r\'{e}my Le Deunff}
\address{Max-Planck-Institut f\"ur Physik komplexer Systeme,
  N\"othnitzer Str. 38, D01187 Dresden, Germany}

\author{Muhan Choi}
\email{mhchoi@ee.knu.ac.kr}
\address{School of Electronics Engineering, Kyungpook National University, Daegu, 702-701, South Korea}

\author{Roland Ketzmerick}
\address{Max-Planck-Institut f\"ur Physik komplexer Systeme,
  N\"othnitzer Str. 38, D01187 Dresden, Germany}
\address{Technische Universit\"{a}t Dresden, Institut f\"{u}r Theoretische Physik
and Center for Dynamics,
 01062 Dresden, Germany}

\date{\today}


\begin{abstract}
We study the quantum Goos-H\"{a}nchen (GH) shift and the tunneling transmission at a curved step
potential by investigating the time evolution of a wave packet.
An initial wave packet is expanded in terms of the eigenmodes of a circular step potential. Its time evolution
is then given by the interference of their simple eigenmode oscillations.
We show that the GH shift along the step boundary can be explained by the
energy-dependent phase loss upon reflection, which is defined by modifying
the one-dimensional (1D) effective potential derived from the 2D circular system.
We also demonstrate that the tunneling transmission of the wave packet is
characterized by a free-space image distant from the boundary. The tunneling transmission exhibits
a rather wide angle divergence and the direction of maximum tunneling is slightly rotated from the
tangent at the incident point, which is consistent with the time delay
of the tunneling wave packet computed in the 1D modified effective potential.
\end{abstract}


\maketitle

\section{Introduction}
When an optical beam is totally reflected at a dielectric interface,
the reflected beam comes out at a point shifted along the interface from the incident point. This deviation is called Goos-H\"{a}nchen (GH) shift \cite{GH47}.
This arises due to the interference of its plane wave components
whose phase loss upon the reflection varies with their incident angle.
Although it has been first observed in optical beams,
the GH shift (or effect) is a generic phenomenon of wave mechanics and
can be found in various physical systems such as
graphene \cite{Beenakker09,Sharma11}, normal/superconductor interface \cite{Lee13},
optical beam \cite{Loffler12,Gotte12}, dielectric microcavity
\cite{Lee05,Hentschel02,Schomerus06} or in the dynamics of neutron \cite{Haan10}.

Recently, the GH effect has been studied with electron waves in condensed matters,
$p$/$n$-doped interface in graphene \cite{Beenakker09} and
normal/superconductor (NS) interface \cite{Lee13}.
Note that these works have investigated the GH effect appearing at a {\it planar} interface,
where it is not difficult to find the phase loss, occurring upon reflection,
for a plane wave component.
However, if we deal with a {\it curved} interface, the plane wave decomposition
becomes useless as the plane wave components are no longer satisfying the matching conditions
at the curved boundary.
Thus a straightforward extension of the method used in the planar interface
cannot be applied to understand the GH shift appearing along the curved boundary
and it becomes necessary to provide a qualitative/quantitative description in such a case.
The GH shift at a curved interface has been
studied by a few authors and only for optical systems.
The contribution of the GH shift in a resonance mode of a circular
microcavity has been estimated by comparing the decay rate expected from a multiple reflection
with the eigenvalue of a resonance mode \cite{Hentschel02}.
Also, the deviation of an optical-beam path from the specular ray reflection
has been discussed when an incident beam comes from outside the cavity and reaches
the circular interface \cite{Schomerus06}.

In addition to the GH effect, the curvature of the boundary induces tunneling
transmission which cannot be observed with the simple planar interface.
It will occur at a range of energy just higher than the height of the potential and
corresponds to {\it dynamical tunneling} \cite{Davis81}, even though it can be
interpreted as tunneling through the barrier of an effective 1D potential,
as we shall see later on.
It is noted that both the GH shift and the tunneling arise from the intrinsic nature
of (quantum) waves and have no classical correspondence.
Recently, considerable efforts have been made to understand tunneling phenomena
based on the associated classical dynamics in various systems
such as quantum map \cite{Lock10,Backer10,Mertig13,Shudo12},
optical microcavities \cite{Creagh07,Yang10,Shinohara10,Lee11}
or quantum potentials \cite{Creagh02,Deunff10}.

In this paper, we study the time evolution of wave packets in the 2D
circular step potential. We investigate the quantum GH shift and the tunneling
transmission observed in the wave packet dynamics. Using the eigenmodes
(bound modes and resonance modes) of the system introduced in Section~\ref{sec:eigenmodes},
we construct an initial
wave packet located deep inside the circular cavity, and its subsequent time evolution
is then described by the interference of simple oscillations in time of individual
eigenmodes (Sec.~\ref{sec:expansion}). Section~\ref{sec:gh_shift} is dedicated to the GH shift. We evaluate the phase
change upon reflection by modifying partly the effective potential derived from the 2D system. It follows from the assumption that the GH shift is relevant only to the barrier region of the potential.
It is shown that the variation of the phase loss with respect to the energy is proportional to
 a time delay along the radial direction and is found to be consistent with the GH shift
observed numerically.
In Section~\ref{sec:tunneling}, we demonstrate that the transmitted (tunneling) wave packet
exhibits a free-space image with a distance $\Delta$ from the circular boundary.
The distance $\Delta$ can be understood as an average over individual
distances $\Delta_{j}$ given by the angular momentum conservation in the $j$-th resonance mode. Then, we shall provide a qualitative explanation of the transmitted angle $\alpha_T$
of the maximum tunneling direction in terms of the time delay of a 1D wave packet in a simplified 1D effective potential. Finally, we shall discuss the case of high-energy wave packets in Section~\ref{sec:high_energy}.

\section{Eigenmodes: bound and resonance modes}\label{sec:eigenmodes}
\begin{figure}[tbp]
  \centerline{
\includegraphics[width=9cm]{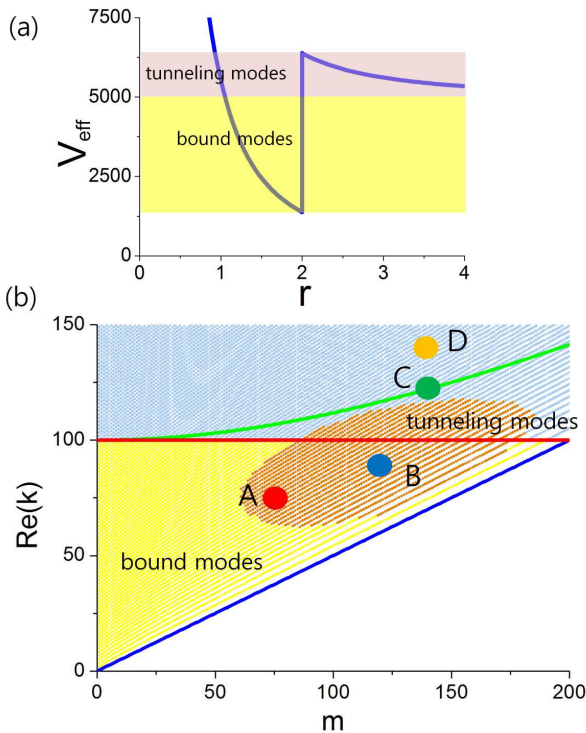}
  }
  \caption{(Color online) (a) The effective potential $V_\text{eff} (r)$ defined by Eq.~(\ref{pot_eff}) for $m=120$ and $V_0=5000$ is depicted by the blue line. The yellow and purple color correspond to the energy ranges where the bound and tunneling modes lie, respectively. (b) For the same $V_0$, the eigenmodes are represented in the $(m,\mbox{Re}(k))$ plane. The central values of $m$ and $k$ of four wave packets (A-D) are marked by large solid dots, i.e. $(m_0,k_0)=(75,75)$ (red), $(120,90)$ (blue), $(140,122.065)$ (green), and $(140,140)$ (yellow), respectively. The tiny orange dots around B mark the relevant eigenmodes used to reconstruct the wave packet B (see Sec.~\ref{sec:expansion}).}
  \label{modes}
\end{figure}

We consider a 2D circular step potential with zero potential inside
the circle and $V_0$ outside (see Fig.~\ref{expand}~(a)), i.e.
\[
V(r)=\left\{
  \begin{array}{l l}
    0 & \quad \text{if} \,\,\, r< R \;; \\
    V_0 & \quad \text{if} \,\,\, r \ge R \;,\\
  \end{array} \right.
\]
where $R$ is the radius of the circle. The rotational symmetry allows us to separate
the time-independent Schr\"odinger equation
\begin{equation}
\left(-\frac{\hbar^2}{2m^*}\nabla^2 + V(r)\right)\Psi(r,\theta) = E\Psi(r,\theta)\;,
\end{equation}
into an angular and a radial equation by writing the wavefunctions $\Psi(r,\theta)$ as a product of the angular and radial solutions, respectively $\Theta(\theta)$ and $\Phi(r)$.
The solutions $\Theta(\theta)$ of the angular part can be expressed as
\begin{equation}
\Theta (\theta) \sim \cos m\theta,\,\, \sin m\theta,
\end{equation}
where $m$ is a natural integer corresponding to the angular momentum quantum number.
The radial equation is written as
\begin{equation}
-\frac{\hbar^2}{2m^*} \left[ \frac{\partial^2 }{\partial r^2} +
\frac{1}{r}\frac{\partial}{\partial r} \right] \Phi(r)+ V_\text{eff} (r)  \Phi(r)= E \Phi(r),
\label{radialeq}
\end{equation}
where $m^*$ is the particle mass, and the effective 1D potential is defined as
\begin{equation}\label{pot_eff}
V_\text{eff} (r)= \left\{
  \begin{array}{l l}
    \frac{\hbar^2 m^2}{2m^*r^2} & \quad \text{if} \,\,\, r< R \;; \\
    V_0 +\frac{\hbar^2 m^2}{2m^*r^2}& \quad \text{if} \,\,\, r \ge R\;.\\
  \end{array} \right.
\end{equation}
For the sake of clarity, we will take $m^*=\hbar=1$, and $R=2$ and $V_0=5000$ for the numerical calculations throughout the paper, unless other values are explicitly specified.
An example of the effective potential $V_\text{eff} (r)$ with $m=120$ is shown in Fig.~\ref{modes}~(a).
It is easy to see that, in addition to the low energy range ($E < V_0$) associated with the bound modes, there is an energy range in which tunneling transmission can take place,
i.e. $V_0 <E < V_0 + \hbar^2 m^2/2m^*R^2$.

The radial equation is nothing but the Bessel equation and  the
solution inside the circle is the Bessel function
\begin{equation}
\Phi_I(r) \sim J_m(k r) \quad \text{if}  \,\,\, r \le R\;,
\end{equation}
where $k \equiv \sqrt{(2m^*/\hbar^2)E}$.
Outside the circular cavity, the solution must be a decaying function
when the energy $E$ is less than the step height $V_0$, while it
would be a propagating function for $E>V_0$.
These solutions correspond to the modified Bessel function and the
Hankel function, respectively
\begin{equation}
\Phi_{II}(r) \sim \left\{
  \begin{array}{l l}
    K_m(\kappa r)& \quad \text{if} \,\,\, r \ge R,\quad E< V_0 \;;  \\
    H^{(1)}_m(k_\text{out} r) & \quad \text{if} \,\,\, r \ge R, \quad E>V_0\;,\\
  \end{array} \right.
\end{equation}
where  $\kappa \equiv \sqrt{(2m^*/\hbar^2)(V_0-E)}$ and $k_\text{out} \equiv \sqrt{(2m^*/\hbar^2)(E-V_0)}$.
When $E>V_0$ we impose the outgoing-wave boundary condition by taking only  the Hankel function of the first kind
 $H^{(1)}_m(k_\text{out} r)$. Therefore, the resonance modes shall decay in time and thus are characterized by the decay rate $\gamma$ corresponding to the imaginary part of the complex energy,
$E=\omega -i\gamma/2$. The complex energy indicates that we are dealing with a non-Hermitian system where the eigenmodes are not orthogonal to each other, as we shall discuss in the next section.

The solutions $k_{mn}$ or $E_{mn}(=\hbar^2 k_{mn}^2/2m^*)$ are extracted by
imposing the boundary conditions,
\[
\Phi_I(R)=\Phi_{II} (R), \quad \frac{d\Phi_I}{dr}(R)=\frac{d\Phi_{II}}{dr} (R)\;.
\]
Since our system is integrable, we have two good quantum numbers $(m,n)$
specifying all the eigenmodes, where $m$ and $n$ are the angular momentum and radial quantum numbers,
respectively. The $k_{mn}$'s are shown in the $(m,\mbox{Re}(k))$ plane in Fig.~\ref{modes}~(b).
The blue line denotes $k_B=m/R$ corresponding to the bottom of the effective potential.
The red line is $k_V=\sqrt{(2m^*/\hbar^2)V_0}$\ and separates the bound and resonance modes.
Among the resonance modes, the modes below the green line,
$k_T=\sqrt{(2m^*/\hbar^2)V_0+ m^2/R^2}$, are the tunneling modes, decaying via a tunneling process, while the modes above are named leaky modes.
The large solid dots A,B,C,D denote the central values $(m_0,k_0)$
of various wave packets discussed in the next sections.

\section{Inner product and Eigenmode expansion}\label{sec:expansion}
\begin{figure}
  \centerline{
 \includegraphics[width=9cm]{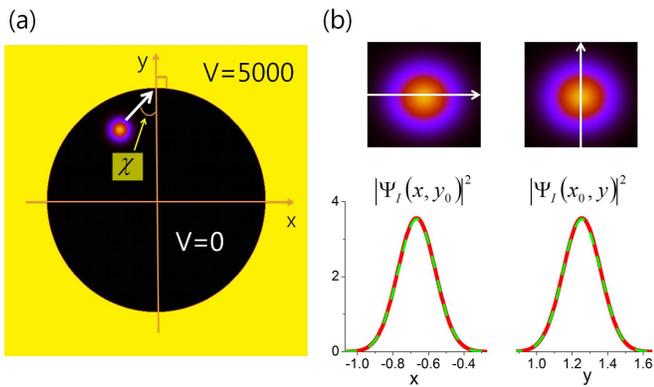}
  }
  \caption{(Color online) (a) The initial Gaussian wave packet $|\Psi_I(x,y)|^2$
with $(m_0,k_0)=(120,90)$ is plotted with the 2D potential in the ($x,y$)-plane. (b) Comparison between the original Gaussian wave packet defined by Eq.~(\ref{gaussfree}), depicted by the red line, and the reconstructed wave packet (dashed green line) obtained from Eq.~(\ref{gaussexp}).
The  horizontal and vertical profiles of $|\Psi_I(x,y)|^2$ are shown.
The center of $|\Psi_I(x,y)|^2$ is located at $(x_0,y_0)$.}
\label{expand}
\end{figure}

It is well known that the bound modes satisfy the orthogonality relation,
\begin{equation}
\langle \tilde{\phi}_i | \tilde{\phi}_{i'} \rangle =\delta_{ii'}\;,
\end{equation}
where the tilde indicates that
the mode is normalized as follows
\begin{equation}
|\tilde{\phi}_i\rangle=\frac{1}{\sqrt{\langle \phi_i | \phi_i \rangle}}
|\phi_i\rangle\;.
\end{equation}
The inner product is defined in the position representation as
\begin{equation}
\langle \phi_i | \phi_i \rangle =\int d\vec{x} \, \phi^*_i(\vec{x}) \phi_i(\vec{x})\;,
\end{equation}
where the star stands for the complex conjugate of the wavefunctions and $\vec{x}$ is a vector in the coordinate space, the $(x,y)$-plane in our case.

However, the resonance modes $|\phi_j\rangle$ are neither square integrable
nor orthogonal, but hold the  biorthogonality relation between the left $\langle \tilde{\psi}_j |$
and the right eigenmodes $|\tilde{\phi}_j\rangle$,
\begin{equation}
\langle \tilde{\psi}_j | \tilde{\phi}_{j'} \rangle =\delta_{jj'}\;,
\end{equation}
where
\begin{eqnarray}
|\tilde{\phi}_j\rangle &=& \frac{1}{\sqrt{\langle \psi_j | \phi_j \rangle}} |\phi_j\rangle \;;  \\
\langle \tilde{\psi}_j | &=& \frac{1}{\sqrt{\langle \psi_j | \phi_j \rangle}} \langle \psi_j |\;.
\end{eqnarray}
We shall consider systems with time-reversal symmetry where
the left eigenmode  $\langle \psi_j|$ is merely the complex conjugate
of the right eigenmode \cite{Lee09}, and the inner product is thus written as
\begin{equation}
\langle \psi_j | \phi_j \rangle=\langle \phi_j^* | \phi_j \rangle=
\int d\vec{x}\,\, \phi_j^2 (\vec{x})\;.
\end{equation}

For the circular case, we can derive analytic expressions for the above inner products.
With $Z_m(x) \equiv J_m(x)$ or $H^{(1)}_m (x)$, one can use the indefinite-integral relation
\[
\int dx \, x \, Z_m^2 (x) = \frac{1}{2} x^2 \bigg[ Z_m^2(x) - Z_{m-1} (x) Z_{m+1} (x)\bigg] \;,
\]
to obtain the inner product of the resonance modes
\begin{eqnarray}
\langle \phi^*_j |\phi_j \rangle &=& \frac{\pi R^2}{2} \bigg[ J_m^2(z)-J_{m-1}(z)J_{m+1} (z)\bigg]  \\
  & &-\frac{c^2 \pi R^2}{2} \bigg[ H_m^{(1)2}(z')-H_{m-1}^{(1)}(z')H_{m+1}^{(1)} (z') \bigg]\;,\nonumber
\end{eqnarray}
where the complex arguments are $z \equiv k R$ and $z' \equiv k_\text{out} R$, and $c \equiv H_m^{(1)}(z')/J_m(z)$.
A similar expression can be derived for the bound modes, replacing $H^{(1)}_m (x)$ by $K_m(x)$,
\begin{eqnarray}
\langle \phi_i |\phi_i \rangle &=& \frac{\pi R^2}{2} \bigg[ J_m^2(z)-J_{m-1}(z)J_{m+1} (z)\bigg]  \\
  & &-\frac{c'^2 \pi R^2}{2} \bigg[ K_m^{2}(z')-K_{m-1}(z')K_{m+1} (z') \bigg]\;,\nonumber
\end{eqnarray}
where $z \equiv k R$ and $z' \equiv \kappa R$, and $c \equiv K_m(z')/J_m(z)$. Note that $k$, $\kappa$, and $c$  are real in the latter case.

Assuming the completeness relation
\begin{equation}
\sum_i |\tilde{\phi}_i\rangle \langle \tilde{\phi}_i | +
\sum_j|\tilde{\psi}_j\rangle \langle \tilde{\phi}_j | =1\;,
\end{equation}
where $i$ and $j$ run over the bound and resonance modes, respectively,
an arbitrary state $|\Psi \rangle$ can be expanded in terms of the eigenmodes of our system
as follows
\begin{equation}
|\Psi \rangle = \sum_i \langle \tilde{ \phi}_i | \Psi \rangle |\tilde{\phi}_i\rangle
+ \sum_j \langle \tilde{\psi}_j | \Psi \rangle |\tilde{\phi}_j \rangle\;.
\end{equation}

We consider now a 2D Gaussian wave packet
with an average momentum $\vec{k}_0$,
\begin{equation}
\Psi_G (\vec{x},t_0)=\frac{\sqrt{\sigma/\pi}}{1+\imat \sigma t_0}
 e^{\frac{-(\sigma/2)|\vec{x}-\vec{x}_C|^2+\imat \vec{k}_0\cdot (\vec{x}-\vec{x}_C)-\imat E_0 t_0/\hbar }
{1+\imat\sigma t_0 } } \;,
\label{gaussfree}
\end{equation}
where $E_0=|\hbar k_0|^2/2m^*$.
This is an exact solution of the free-particle time-dependent
Schr\"{o}dinger equation.
The numerical calculations shall be carried out for $\sigma=100$
and the position of the wave packet $\vec{x}_C=(0,R)$ at $t_0=0$.
 We choose as an initial wave packet the Gaussian wave packet
at $t_0=-1/k_0$, located deep inside the cavity, a unit length apart from the $\vec{x}_C=(0,R)$
(see Fig.~\ref{expand}~(a)).
In order to avoid any ambiguities in the determination of their positions and velocities, the parameters of all the considered wave packets will always be chosen such that their typical sizes are much smaller than the radius $R$ of the potential.
This initial wave packet would hit the point $\vec{x}_C=(0,R)$ at the boundary after a time interval $s_0=1/k_0$. For the following, we will conveniently set the time unit as $s_0$.

Assuming a Gaussian wave packet with an angular momentum $m_0\hbar$ and a central wave number $k_0$,
as indicated by the large solid dots in Fig.~\ref{modes}~(b), then its central energy is
$\hbar^2 k_0^2/2m^*$ and the incident angle $\chi$ (see Fig.~\ref{expand}~(a))
is determined by the semiclassical relation $\sin \chi=m_0/k_0R$. The mean position of the initial wave packet
$\Psi_I (\vec{x})=\Psi_G (\vec{x},t_0=-1/k_0)$
is given by $(x_0,y_0)=(-m_0/k_0R, R-\sqrt{1-(m_0/k_0R)^2})$.

The initial state can be expanded as
\begin{equation}
|\Psi_I \rangle = \sum_i a_i |\tilde{\phi}_i\rangle
+ \sum_j b_j  |\tilde{\phi}_j \rangle\;,
\label{gaussexp}
\end{equation}
where the overlap of the initial wave packet with the bound and resonance modes are respectively given by $a_i \equiv \langle \tilde{ \phi}_i | \Psi_I \rangle$ and $b_j \equiv \langle \tilde{\psi}_j | \Psi_I \rangle $.
For instance, the reconstruction of the initial wave packet B, with $(m_0,k_0)=(120,90)$ (see Fig. 2 (a)), has been done using the expansion Eq.~(\ref{gaussexp}) over 1982 eigenmodes (1374 bound modes and 608 resonance modes, corresponding to the tiny orange dots in Fig.~\ref{modes}~(b)), and presents an excellent agreement with the gaussian wave packet (Eq.~(\ref{gaussfree})) as shown in Fig.~\ref{expand}~(b).
It turns out that the reconstruction of the initial wave packet by Eq.~(\ref{gaussexp})
works well for the low-energy wave packets (A, B), but presents some discrepancies for the high-energy wave packets (C,D), as shown in Fig.~\ref{criticalwp}~(a) and (b) where some modulation of the incident wave packets is seen.

\section{GH shift in wave packet dynamics}\label{sec:gh_shift}

\begin{figure}
  \centerline{
 \includegraphics[width=8cm]{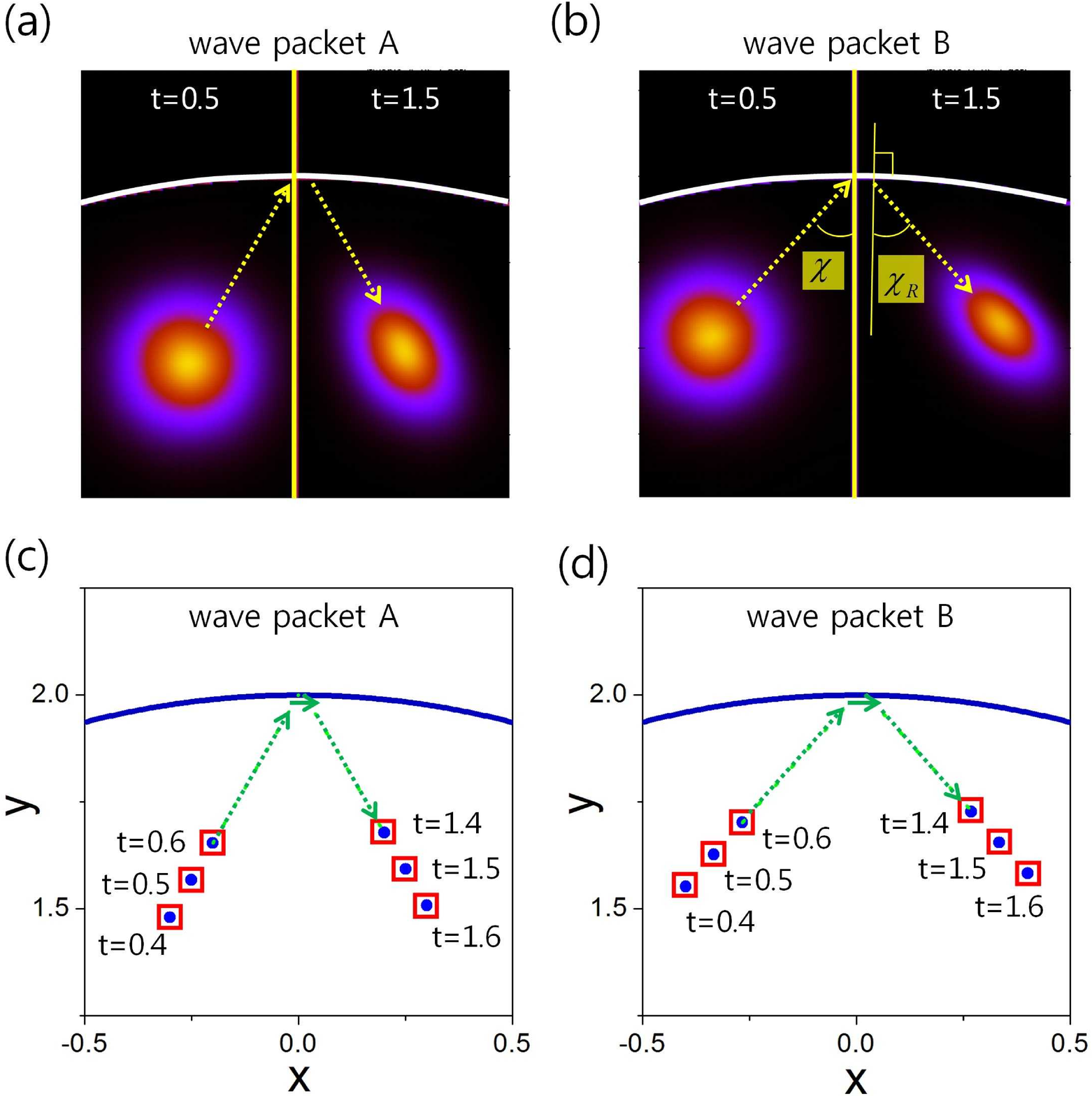}
  }
  \caption{(Color online) The time evolution of the wave packets is shown in the $(x,y)$-plane. The wave packets A and B are plotted in (a) and (b), respectively, at $t=0.5$ and $1.5$. The thick vertical yellow line corresponds to the y-axis while the white curved line depicts the boundary of the circular cavity. In the panel (c) and (d), the red rectangles are the expected positions of the wave packets A and B at various times obtained from Eq.~(\ref{GH_spat_shift}) with 1 \% and 2.5 \% increased reflection angles, respectively, and show a very good agreement with the average positions $\vec{x}_{av}(t)$ denoted by the solid blue dots.}
  \label{ghshift}
\end{figure}

\subsection{Numerical results}
The time evolution of the initial wave packet is obtained by integrating the time-dependent Schr\"odinger equation and leads straightforwardly to
\begin{equation}
|\Psi (t) \rangle = \sum_i a_i e^{-\imat E_i t/\hbar} |\tilde{\phi}_i\rangle
+ \sum_j b_j e^{-\imat E_j t/\hbar} |\tilde{\phi}_j \rangle\;.
\label{wp_expand}
\end{equation}
The wave packet would hit the boundary point $(0,R)$ at $t=1$ in the time unit of $s_0=1/k_0$.
The previous expression is nothing but Eq.~(\ref{gaussexp}) with the dynamical phase factors containing the eigenvalues $E_i$ and $E_j$.

As an example, let us first consider the wave packet A with $(m_0,k_0)=(75,75)$, which
has an incident angle $\chi= \pi/6$ and an energy $E_0=(\hbar k_0)^2/2m^*=2812.5$ far below $V_0=5000$.
We expand the initial wave packet in terms of 2474 eigenmodes (2428 bound modes and
66 resonance modes), which corresponds to the modes with an overlap $|a_i|$ (or $|b_i|$)
greater than 0.0005. The wave packet A at $t=0.5$ and $t=1.5$ is shown in Fig.~\ref{ghshift}~(a).
The green dotted arrows indicate the path of the top of the wave packet and the GH shift is already clearly discernible. It is numerically evaluated at $l_{GH}\simeq 0.015$.

For the wave packet B with $(m_0,k_0)=(120,90)$
and $\chi \simeq 0.73$ or $41.8$ degrees,
 the same picture is shown in Fig.~\ref{ghshift}~(b).
In this case, the GH shift is $l_{GH}\simeq 0.024$.
Although the energy  $E_0=4050$ is less than $V_0$,
it contains many tunneling modes (608 tunneling modes) in addition to the bound modes (1374 bound modes)
as shown in Fig.~\ref{modes}~(b), yielding an appreciable tunneling transmission
that we shall discuss later on.

The shift $l_{GH}$ is computed using a more systematic evaluation of the wave packet position,
by calculating the average position for a given $t$,
\begin{equation}
\vec{x}_{av}(t) =\frac{\int d\vec{x}\,\, \vec{x} \, |\Psi (\vec{x},t)|^2}
{\int d\vec{x} \, |\Psi (\vec{x},t)|^2},
\end{equation}
when the wave packet is sufficiently far away from the reflection point $\vec{x}_C=(0,R)$.
The wave packet positions at $t=0.4,0.5,0.6$ and $t=1.4,1.5,1.6$ are shown
(blue solid dots) in Fig.~\ref{ghshift}~(c) and (d) for the wave packets A and B, respectively.

\subsection{Theory for the GH shift at a curved boundary}

\begin{figure}
  \centerline{
 \includegraphics[width=8cm]{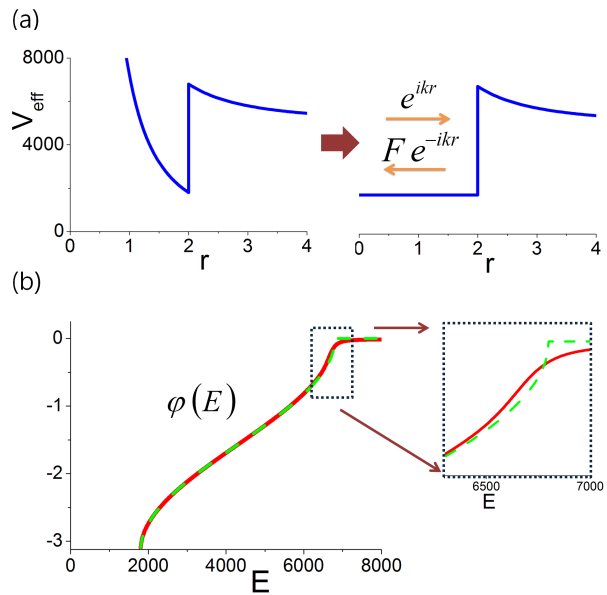}
  }
  \caption{(Color online) The effective potential $V_\text{eff}(r)$ (Eq.~(\ref{pot_eff})) is modified according to the panel (a), from left to right, in order to define the phase change $\varphi_R$ upon reflection. The effective potential is plotted for $m=120$ and $V_0=5000$. On the panel (b) the phase change $\varphi_R$ (red solid line) is plotted with respect to the energy and compared to the phase $\phi$ of the 1D step potential case (see Eq.~(\ref{stepphase})) with ${\cal V}_0=5000$. The phases are nearly the same except for higher energy as shown in the inset.}
  \label{modelgh}
\end{figure}

In this section, we provide a simple explanation of the GH shift observed in the wave packet dynamics.
In order to gain some insight into the theory, it is illuminating to mention first
the simple case of the {\it planar} step potential.
For a 1D step potential with a height ${\cal V}_0$, the reflection coefficient
is expressed as $F=e^{\imat \phi}$ when the energy ${\cal E}$ of the incident plane wave
is less than ${\cal V}_0$.
The phase of the reflection coefficient has an analytic expression
\begin{equation}
\phi=2\arctan \sqrt{\frac{{\cal E}}{{\cal V}_0-{\cal E}}} -\pi.
\label{stepphase}
\end{equation}
Then the wave packet dynamics exhibits a time delay upon reflection \cite{Bracher95,Lee13}
which is given by the change of the phase $\phi$ with respect to ${\cal E}$,
\begin{equation}
\delta t_R = \hbar \frac{d\phi}{d{\cal E}}=
\frac{\hbar}{{\cal V}_0} \left( \sqrt{\frac{{\cal V}_0-{\cal E}}{{\cal E}}}
+ \sqrt{\frac{{\cal E}}{{\cal V}_0-{\cal E}}} \right),
\label{tR}
\end{equation}
and it is also known as the Wigner time scale \cite{Wigner55}.
Therefore, it is straightforward to see that the GH shift in the 2D planar case is characterized by a spatial shift $l_{GH}$ which is merely related to the time delay $\delta t_R$ by the expression
\begin{equation}
l_{GH}=v_y \delta t_R,
\label{GH_spat_shift}
\end{equation}
where $v_y$ is the velocity component parallel to the step boundary.

Returning to the circular case, let's take advantage of the 1D effective potential $V_\text{eff}(r)$ from which we want to extract the phase loss
$\varphi_R$ under reflection in the same spirit as in the planar case.
To do this, we assume that the time delay $\delta t_R$ is only determined
by the shape of the barrier itself and the form of the potential where the incident wave packet evolves before hitting this barrier is not relevant. We can then modify the left part of potential
into a constant potential so that we can consider plane wave solutions and the reflection
coefficient $F=|F|e^{\imat \varphi_R}$ in this region as illustrated in Fig.~\ref{modelgh}~(a).
The reflection coefficient $F$ is determined by matching the solutions
at the frontier between the free region, i.e. the plane waves,
and the barrier region, i.e. $K_m (\kappa r)$ for $E<V_0$ and $H^{(1)}_m (k_\text{out} r)$ for $E>V_0$.
The figure \ref{modelgh}~(b) shows the resulting phase loss $\varphi_R$ (solid red line) which is very similar to the phase loss $\phi$ of the step potential
given by Eq.~(\ref{stepphase}), except for a slight deviation
in the high energy range of $V_0 < E < V_0+(\hbar m)^2/2m^*R^2$
(see the dashed green line in the inset).
Then, similarly to the 2D planar case, defining the time delay in the radial direction as
$\delta t_R= \hbar \frac{d\varphi_R}{d E}$,
the GH shift is given by
\begin{equation}
l_{GH} = v_{\theta} \delta t_R= (\frac{m_0 \hbar}{R})\hbar \frac{d\varphi_R}{dE}\;,
\label{ghtheory}
\end{equation}
where $v_\theta$ is the velocity of the wave packet projected along the $\theta$-direction.

For the wave packet A with $(m_0,k_0)=(75,75)$, Eq.~(\ref{ghtheory}) gives $l_{GH}=0.0152$ and $t_R \simeq 4.05 \times 10^{-4} \simeq 0.0304 s_0$.
The resulting trajectory is depicted by the red rectangles in Fig.~\ref{ghshift}~(c) where we used, for a better fit, a slightly increased reflection angle $\chi_R=1.01\chi$ compared to the incident angle. This leads to an excellent agreement with the motion $\vec{x}_{av}(t)$ of the wave packet computed numerically, illustrated by the blue dots on Fig.~\ref{ghshift}~(c).
Similarly, for the wave packet B with $(m_0,k_0)=(120,90)$ we  get
$l_{GH}=0.0242$
and $t_R \simeq 4.05 \times 10^{-4} \simeq 0.0363 s_0$ from Eq.~(\ref{ghtheory}). The reflection angle is also slightly modified $\chi_R=1.025\chi$ and
we see again a good agreement between the theory and the numerical results from the wave packet evolution as shown on Fig.~\ref{ghshift}~(d).
The correction in the fit of the reflected angles can be justified as follows: on contrary to the 2D step potential case, in the circular cavity, the reconstructed wave packets (for instance A and B) contain tunneling-mode components which are responsible for a slight increase of the reflected angle.
The reduction of smaller-incident-angle components, due to tunneling transmission,
in the reflected wave packet results in the slight increase of the reflection angle $\chi_R$.
In optics, this is called ``Fresnel filtering effect'' \cite{Rex02}.
For a generic continuous 2D potential barrier, the GH shift would appear in a rather complex way due to the mixing with the change of dynamical phase of the rounded reflection at the continuous barrier.

\section{Tunneling transmission}\label{sec:tunneling}

\begin{figure}
  \centerline{
 \includegraphics[width=8cm]{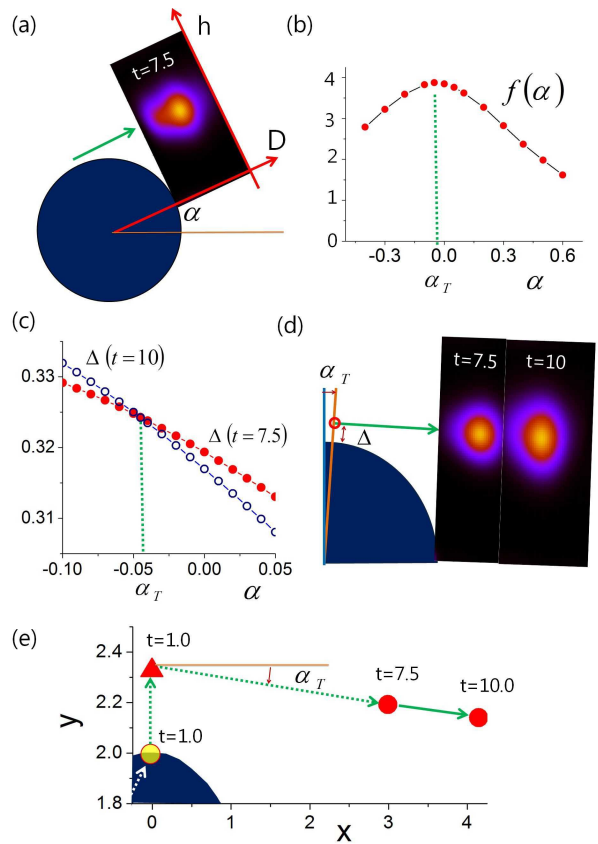}
  }
  \caption{(Color online) Tunneling transmission of the wave packet B.
  (a) Schematic illustration of the coordinates $(D,h)$ and the rotation angle $\alpha$ for the emission Husimi function $H_E(D,h;\alpha)$. (b) The strength of tunneling transmission $f(\alpha)$ defined by Eq.~(\ref{f_alpha}) is evaluated at $t=7.5$ (red dotted line). The vertical green dotted line indicates the maximum of $f(\alpha)$. (c) The distance $\Delta$ of the normal tunneling wave component
 is plotted as a function of the angle $\alpha$ at two different times $t=7.5$ and $t=10$ (respectively the red solid dots and the open blue circles). The vertical green dotted line marks the value $\alpha_T=-0.045$ where the curves intersect.
  (d) The normal tunneling wave component
   is plotted at $t=7.5$ and $t=10$ for the particular value $\alpha_T=-0.045$. (e) The average positions $\vec{x}_{av}$ of the normal tunneling wave component
     are shown in the $(x,y)$-plane. The position at $t=1$ (solid red triangle) is extrapolated from the observed positions at $t=7.5$ and $t=10$ (solid red circles).}
  \label{tunnelwp}
\end{figure}

\subsection{Numerical results}
As mentioned previously, the wave packet B contains many tunneling-mode components.
So, we can expect to detect a transmitted wave packet outside of the circular region.
In optical microcavities, the tunneling transmission has been characterized by its free-space image with a distance $\Delta$ from the cavity surface \cite{Creagh07,Lee11,Tomes09}.
The characterization by the distance $\Delta$ is also pertinent in our case
since both transmissions occur through the same mechanism, the
tunneling process.

In order to determine the most probable direction of the tunneling transmission, we use
the emission Husimi function used in Ref. \cite{Lee11}, which is modified to
be a wave-number independent function in the present case.
We consider a projection line ($h$-axis) which has been rotated
by an angle $\alpha$ and is a distance $D$ apart from the origin
as shown in Fig.~\ref{tunnelwp}~(a). The modified emission Husimi function
on the projection line is defined as
\begin{equation}
H_E (D,h;\alpha) = \left| \int dh' \, \Psi(x(h'),y(h'))\xi(h,h') \right|^2
\label{husimi_em}
\end{equation}
where $\Psi(x,y)$ is the 2D wave
function at a given time defined in Eq.~(\ref{wp_expand}) in position representation, $h$ is the coordinate on the line and
$\xi(h,h')=\frac{1}{(\mu \pi)^{1/4}} e^{-(h'-h)^2/2\mu}$ with $\mu=(0.15)^2/2$.
This emission Husimi function probes the intensity of
the wave component normally incident to the projection line.
Any point $(x,y)$ is related to $(D,h)$ at an angle $\alpha$ as
\begin{eqnarray}
x&=&D \cos \alpha -h \sin \alpha,  \nonumber \\
y&=&D \sin \alpha + h \cos \alpha.
\end{eqnarray}
For a given value of $\alpha$, we can track the tunneling wave
component, normally incident to the $h$-axis, by plotting
$H_E(D,h)$, as shown in Fig.~\ref{tunnelwp}~(a) and (d).
We can then obtain the average position $(D_0,h_0)$ of the emission Husimi function
in the $(D,h)$ plane and the strength of the
normal tunneling component defined as
\begin{equation}
f(\alpha )=\iint{dD\, dh\, H_E (D,h;\alpha)}
\label{f_alpha}
\end{equation}
at a given time $t$.
The strength $f(\alpha)$ evaluated at $t=7.5$ is shown
as a function of $\alpha$ in Fig.~\ref{tunnelwp}~(b)
and the plot exhibits a maximum at $\alpha_T\simeq -0.045$.
This angle $\alpha_T$ indicates the most probable
direction of the tunneling transmission (see Fig.~\ref{tunnelwp}~(d)).
This most probable
tunneling direction can be confirmed with the plot of $\Delta=h_0-R$.
The values of $\Delta$  at two distinct times, $t=7.5$ and $t=10$ are shown
as a function of $\alpha$ in Fig.~\ref{tunnelwp}~(c) and it is clear
that we can get the time-independent value $\Delta\simeq 0.324$ from the crossing point of the two curves which corresponds to the tunneling direction $\alpha_T$.
The figure \ref{tunnelwp}~(d) shows the evolution of the normal tunneling wave
component by plotting $H_E(D,h;\alpha_T)$ at $t=7.5$ and $10$. The emission Husimi function
spreads out as the time increases, remaining $\Delta$ invariant.

\begin{figure}
  \centerline{
 \includegraphics[width=6cm]{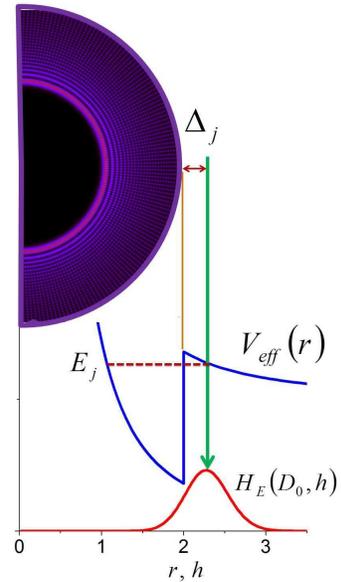}
}
 \caption{(Color online) The resonance mode $(m,k)=(120,113-\imat1.57 \times 10^{-6})$ is shown in the upper part of the picture. In the lower part, the blue line is the corresponding effective potential with $V_0=5000$ and the red line represents the emission Husimi function $H_E (D_0,h)$ of the mode with $D_0=1.5R$, whose maximum gives the individual gap $\Delta_j$. The horizontal dashed line superimposed on the potential denotes the energy $E_j$ of the resonance mode.}
  \label{tunnelmodel}
\end{figure}

The negative value of the angle $\alpha_T$ implies that the tangential tunneling
wave component appears beyond the barrier after some time delay.
The initial position of the tangential tunneling wave
component, i.e. $(x_T,y_T)\simeq ( (R+\Delta)\sin |\alpha_T|, (R+\Delta) \cos \alpha_T )\simeq (0.105,2.321) $, is marked by the open red circle in Fig.~\ref{tunnelwp}~(d).
The transmitted wave component propagates freely and from its average positions,
$(x,y)=(2.993,2.192)$ at $t=7.5$ and $(4.146,2.14)$ at $t=10$,
we can estimate the position $(x,y)=(-0.005,2.326)$ at the time $t=1$.
It is worth noticing that the extrapolated position of the tunneling wave
component at $t=1$
is located at $x\simeq 0$ which is also the position of incident wave packet at $t=1$ as
illustrated in Fig.~\ref{tunnelwp} (e).
We can then estimate the time $t^*$ when the tunneling wave
component is located at $(x_T,y_T)$ and we get $t^*\simeq 1+ 0.105\times (6.5/3.0)\simeq 1.227$.
Therefore, the tunneling wave component
 comes out with a time delay  $\delta t^* \simeq 0.227 s_0$
at the position $(x_T,y_T)$ and  a slightly rotated angle $\alpha_T=-0.045$.

\subsection{Explanation of the distance $\Delta$}

In this section we aim to provide a semiclassical interpretation of the width $\Delta$. We shall illustrate our purpose with help of the wave packet B. In the circular case, each individual resonance with $(m_j,k_j)$
exhibits its own distance $\Delta_j$ and has a clear physical meaning as it corresponds to
the width of the effective potential Eq.~(\ref{pot_eff}) at the given energy $E_j=(\hbar k_j)^2/2m^*$.
This is illustrated in Fig.~\ref{tunnelmodel} for the resonance mode
$(m_j,k_j)=(120,113.0-\imat1.57\times 10^{-6})$.
This relation is obvious from the semiclassical angular momentum
conservation,
\begin{equation}
m_j=(R \sin \chi_j)k_j =(R+\Delta_j) k_{\text{out},j}\;,
\end{equation}
where $k_{\text{out},j} = \sqrt{(2m^*/\hbar^2)(E_j-V_0)}$.
Then, the distance $\Delta_j$ becomes
\begin{equation}
\Delta_j=m_j/\sqrt{(2m^*/\hbar^2)(E_j-V_0)}-R,
\end{equation}
and it is easy to see that $E_j=V_\text{eff} (R+\Delta_j)$.
The value of the width $\Delta_j$ for the chosen mode $(m_j,k_j)$ is confirmed numerically by evaluating the maximum of the emission Husimi function $H_E (D_0=1.5R,h)$ of the eigenmode, as depicted by the red solid line in Fig.~\ref{tunnelmodel}. Let us note that the emission Husimi function of the
eigenmodes does not depend on the parameter $\alpha$ because of their rotational
symmetry.

Taking into account all these resonance modes, one can reasonably expect
the width $\Delta$ to be predicted by summing the contributions of each
resonance component weighted by an appropriate factor, and thus be expressed by the average value over all the $\Delta_j$'s,
\begin{equation}
\Delta=\frac{\sum_j |b_j|^2\gamma_j \Delta_j}{\sum_j |b_j|^2\gamma_j}\;,
\label{delta_average}
\end{equation}
where $\gamma_j$ is the decay rate of the mode, appearing in the imaginary part of
the energy $E_j=\omega_j-i\gamma_j/2$. Performing the sum over the $608$ relevant resonant modes in the case of the wave packet B, Eq.~(\ref{delta_average}) leads to $\Delta\simeq 0.30$  which is consistent with the previous numerical observation $\Delta\simeq 0.324$.

\subsection{Time delay of tunneling wave packet}
\begin{figure}
  \centerline{
 \includegraphics[width=8.5cm]{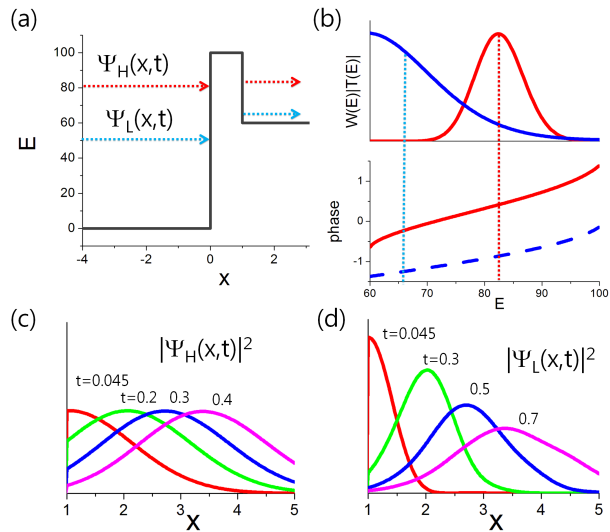}
}
  \caption{(Color online) Dynamics of tunneling wave packets in a rectangular barrier. (a) The potential shape depicted here is given by Eq.~(\ref{pot_rec}) with $V_{\text{max}}=100$, $V_0=60$, $x_a=0$ and $x_b=1$. The red (blue) arrow
approaching the barrier denote the energie of the incident wave packet $\Psi_H$ ($\Psi_L$). The average
energy of tunneling wave packets are marked by the arrows beyond the barrier.
(b) The upper panel describes the weighting distribution $W(E)|T(E)|$ of the plane wave components of the tunneling wave packets $\Psi_H$ and $\Psi_L$ (red and blue line, respectively) as a function
of the energy. The lower panel depicts the phase $\varphi_T$ of the transmission coefficient $T(E)$ (black solid line).
The phase of the reflection coefficient (black dashed line) is also plotted for comparison. The time evolution of the tunneling wave packets $|\Psi_H|^2$ and $|\Psi_L|^2$ is shown in (c) and (d), respectively.
}
\label{stepbarrier}
\end{figure}

In this subsection, we shall provide a qualitative explanation of the rotation
angle $\alpha_T$ of the tunneling wave packet by investigating the time delay of 1D
tunneling wave packets evolving in a 1D effective potential.

First, we start with a simple rectangular barrier as shown in Fig.~\ref{stepbarrier}~(a) and defined as
\begin{equation}
V_{\text{rec}} =
\begin{cases}
 &0 \quad \quad \,\,\, \text{ if } x < x_a\;; \\
 &V_{\text{max}} \quad \text{ if } x_a\leq x\leq x_b\;; \\
 &V_{\text{min}} \quad \, \text{ if } x> x_b\;, \\
\end{cases}
\label{pot_rec}
\end{equation}	
where we shall choose $V_{\text{max}} = 100$, $V_{\text{min}}=60$, $x_a=0$ and $x_b=1$ for the following numerical computations.
We consider an incident wave packet composed of plane wave components,
\begin{equation}
\Psi (x,t)=\int {dk \, W(k,k_0) e^{\imat(k (x-x_a) - E(k)t/\hbar)}}\;,
\end{equation}
where the window function is given by $W(k,k_0)=e^{-(k-k_0)^2/\sigma^2}$ and the central energy of the
incident wave packet is $E_0=(\hbar k_0)^2/2m^*$. This incident wave packet would hit the right side of the barrier ($x=x_a$) at $t=0$. We can get the tunneling
(transmitted) wave packet appearing beyond the barrier by inserting the transmission
coefficient $T(k)$ \cite{Liboff},
\begin{equation}
\Psi (x,t)=\int{dk \, W(k,k_0)T(k) e^{\imat(k'(k) (x-x_b) - E(k)t/\hbar)}}\;,
\label{tunnelwpeq}
\end{equation}
where $k'(k)=\sqrt{(2m^*/\hbar^2)(E(k)-V_{\text{min}})}$. This tunneling wave packet is set to be located
on the right side of the barrier ($x=x_b$) at $t=0$ if $T(k)$ is independent of $k$ (or equally $E$).
However, $T(k)$ depends on $k$ or $E$ in general and it is responsible for the time delay
as discussed below.

\begin{figure}
  \centerline{
 \includegraphics[width=8.5cm]{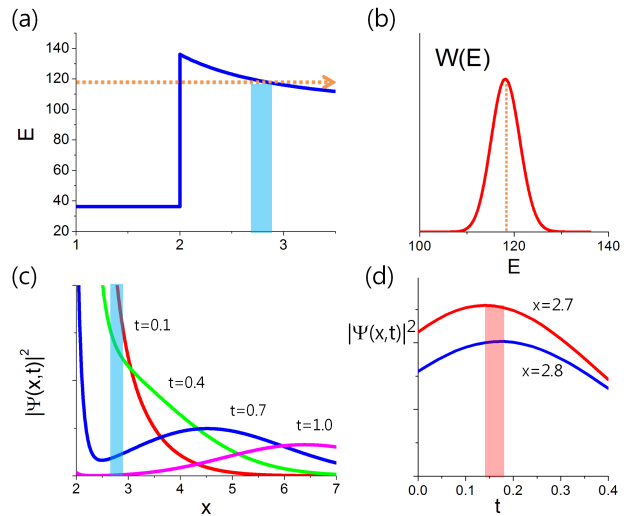}
}
  \caption{(Color online) The dynamics of the tunneling wave packet with an average energy $E_0\simeq 118$ (horizontal orange arrow in (a)) is investigated in the modified effective potential shown in (a) and defined in Fig.~\ref{modelgh} for the parameters $m=17$ and $V_0=100$.
  In the panel (b), the distribution of the plane wave components of the incident wave packet is plotted (red line) with respect to the energy and the maximum of the function is denoted by the vertical orange dotted line. The time evolution of the square modulus of the wave packet is depicted in (c). The time variation of $|\Psi(x,t)|^2$ at $x=2.7$ and $2.8$ is displayed in (d) and the time range delimited by the two maxima (light red window) corresponds to the time delay of tunneling wave packet.}
  \label{modelbarrier}
\end{figure}

First, let us take a narrow window function $W(k,k_0)$ with a small $\sigma$ to construct a wave packet $\Psi_H (x,t)$, choosing $E_0=(\hbar k_0)^2/2m^*=80$ such that its central energy lies in the range $60 < E < 100$ (see Fig.~\ref{stepbarrier}~(a)) where the barrier has a finite but nonzero width. The weighting function $W(k,k_0)T(k)$ of the tunneling wave packet in
Eq.~(\ref{tunnelwpeq}) is shown as a function of $E$ (the red line) in Fig.~\ref{stepbarrier}~(b).
Writing the transmission coefficient as $T(E)=|T(E)|e^{\imat \varphi_T}$ and then performing the stationary-phase approximation on the integral of Eq.~(\ref{tunnelwpeq}), one unravels the semiclassical dynamics of the transmitted (nearly Gaussian-like) wave packet and one extracts, similarly to Eq.(\ref{tR}) for the reflected case, a time delay \cite{Bracher97}
\begin{equation}
\delta t_T =\hbar \frac{d \varphi_T}{dE}\;.
\label{time_delay_trans}
\end{equation}
The phase $\varphi_T$ is shown to increase roughly from $-0.5$ to $1.5$ in the energy range $60 < E < 100$ as depicted by the black solid line in the lower panel of Fig.~\ref{stepbarrier}~(b) and a linear approximation yields the time delay $\delta t_T \simeq 2/40 = 0.05$, which is greater than
the time delay of the reflected wave packet, $\delta t_R \simeq 3/100 = 0.03$ in this case (see the black dashed line in the lower panel of Fig.~\ref{stepbarrier}~(b)).
The evolution of the tunneling wave packet is shown in Fig.~\ref{stepbarrier}~(c).
The time at which the peak of the tunneling wave packet starts to appear at $x_b$ is $t=0.045$ and is consistent with the previous $\delta t_T \simeq 0.05$ estimated from the variation of $\varphi_T$.

Now let us consider another wave packet $\Psi_L (x,t)$ with a lower energy ($E_0=50$) and
a broader window function with a larger $\sigma$. On contrary to $\Psi_H$, the weighting function $W(k,k_0)T(k)$ of this tunneling wave packet
does not have the shape of a Gaussian as shown by the blue line in Fig.~\ref{stepbarrier}~(b).
Nevertheless the time delay, evaluated from the positions of the peak of the tunneling wave packet (see Fig.~\ref{stepbarrier}~(d)), is similar to the previous Gaussian-like packet, i.e., $\delta t_T\simeq 0.045$. The average energies of the transmitted part of $\Psi_H$ and $\Psi_L$, evaluated from the velocity of the peak of the wave packets in Fig.~\ref{stepbarrier}~(c) and Fig.~\ref{stepbarrier}~(d), respectively, are depicted by the red and blue vertical dotted lines on Fig.~\ref{stepbarrier}~(b). One can note that, for both energies, the slope of the phase $\varphi_T$ (lower panel in Fig.~\ref{stepbarrier}~(b)) is very similar and therefore, according to Eq.~(\ref{time_delay_trans}), fully consistent with the comparable time delays $\delta t_T$ observed.

 One can finally obtain the time delay of the transmitted 2D wave packet studied in Sec. V. A
 by considering now the effective potential from Eq.~(\ref{pot_eff}) modified as shown in Fig.~\ref{modelgh}~(a), taking $m=17$ and $V_0=100$ which is nothing but a rescaling of the potential we have been using in Sec.~\ref{sec:gh_shift}~B, since $m^2/V_0=120^2/5000\simeq 17^2/100 \simeq 2.9$. We take a narrow window function $W(k,k_0)$
with $E_0 \simeq 118$ as shown in Figs.~\ref{modelbarrier}~(a,b).
Taking into account the shift due to $T(E)$, the right side of the barrier is
then evaluated at $x_b \simeq 2.7$. The transmitted coefficient $T(E)$ is again
determined by matching the plane wave solutions and the Hankel functions on both
sides of the barrier as already explained in Sec. IV B.
The time evolution of the tunneling wave packet is shown
in Fig.~\ref{modelbarrier}~(c). Once again, the time delay  can be determined by the
maximum point of $|\Psi(x_b,t)|^2$ as illustrated in Fig.~\ref{modelbarrier}~(d).
If we assume $x_b=2.7$, we finally obtain the time delay of $\delta t_T \simeq 0.14$.
Note that this is larger than the reflected time delay $\delta t_R \simeq 3/100=0.03$. Based on the previous calculations for the rectangular barrier, we can also assume the ratio $\delta t_T /\delta t_R \simeq 4.6$ to be
almost constant even when the tunneling wave packet is not Gaussian-like.
Then, one can apply this ratio to our effective potential Eq.~(\ref{pot_eff}) with $m=120$ and
$V_0=5000$. Therefore the expected time delay of the tunneling wave packet in this case
would be $\delta t_T \simeq 4.6 \delta t_R =4.6\times (3/5000) \simeq  0.00276$.
This is consistent with the time delay $\delta t^*=0.227 s_0=0.227/90 \simeq 0.0025$
deduced from the dynamics of the 2D tunneling wave packet in Sec.~\ref{sec:tunneling}~A.


\begin{figure}
  \centerline{
 \includegraphics[width=8cm]{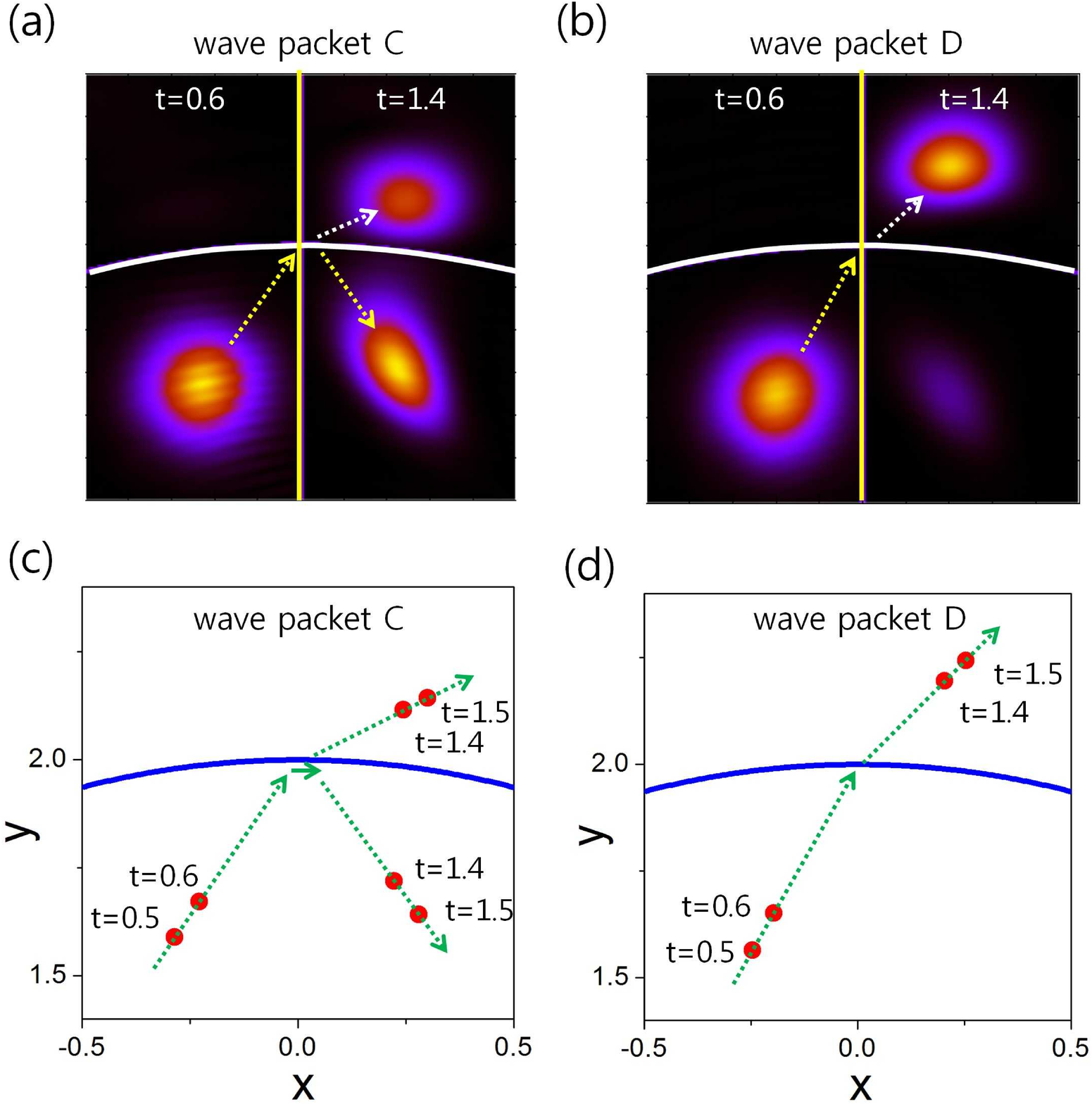}
  }
  \caption{(Color online) Time evolution of the high-energy wave packets. The wave packets C and D are respectively plotted in (a) and (b) in the $(x,y)$-plane at $t=0.6$ and $1.4$ and the average positions $\vec{x}_{av}(t)$ at various times are denoted by red solid dots in (c) and (d) for the wave packet C and D, respectively.
}
  \label{criticalwp}
\end{figure}

\section{High-energy wave packets}\label{sec:high_energy}

Finally, we consider the high-energy wave packets C and D. The central energy of the wave packet C with $(m_0,k_0) \simeq (140,122)$
and $\chi\simeq 35.0$ degrees is chosen to correspond to the top of the barrier of the effective potential
$V_{\text{eff}}$ defined by Eq.~(\ref{pot_eff}) (see Fig.~\ref{modes}~(b)) such that the incident wave packet would be roughly equally distributed into a transmitted and reflected part after the bounce as shown in Fig.~\ref{criticalwp}~(a)
($61.6 \%$ of incident wave packet is reflected).
 Like for the wave packets A and B, the reflected wave packet exhibits a GH shift and an increased reflection angle $\chi_R$ which enhances as the central energy of the wave packet gets larger. Besides, the transmitted wave packet exhibits
a transmission angle of $64.7$ degrees, which is
a considerable deviation from the tangential direction
(the classical expectation of the transmission angle 90.0 degrees) (see Fig.~\ref{criticalwp}~(c)). This corresponds to the Fresnel filtering effect on the transmitted wave packet, well-known in optics \cite{Rex02}.

When the energy becomes even higher like the wave packet D with $(m_0,k_0) \simeq (140,140)$
and $\chi\simeq 30.0$ degrees,
almost the whole wave packet
(about 91.6 \%) would be transmitted as shown in Fig.~\ref{criticalwp}~(b) and (d).
In this deep semiclassical regime, the wave packet follows the classical trajectory of a particle without any appreciable lateral shift.

\section{Summary}

We have studied the dynamics of wave packets in the 2D circular step potential.
The time evolution of the wave packets has been obtained from  the interference
between the time oscillations of eigenmodes of the system, i.e., bound modes and resonance modes.
At low energy, the wave packets exhibit a GH shift along the circular boundary
when they are reflected by the curved step potential. We have shown that
the GH shift is well explained by the variation of the phase of the reflection coefficient which
is obtained using a modified 1D effective potential deriving from the 2D system.
The good agreement with the numerical simulations justifies our assumption that only the shape of
the barrier region is relevant to describe the time delay due to reflection in a 1D potential. We have also demonstrated that the tunneling wave packet is characterized
by a finite distance $\Delta$ from the curved boundary by investigating its dynamics  with the help of the emission Husimi function. The gap $\Delta$ can be estimated by averaging over each individual resonance contributions.
We believe that this wave packet approach will furnish useful materials in order to investigate the natural connections between classical and quantum mechanics.

\section{Acknowledgments}
This research was supported by Kyungpook National University Research Fund, 2013, and
Basic Science Research Program through the National Research Foundation of Korea(NRF) funded by the Ministry of Education(No. 2013R1A1A2065357).
We also acknowledge support by the
Deutsche Forschungsgemeinschaft within the Forschergruppe
760 ``Scattering Systems with Complex Dynamics.''


\end{document}